\documentclass[fleqn,usenatbib]{mnras}

\usepackage{newtxtext,newtxmath}

\usepackage[T1]{fontenc}

\DeclareRobustCommand{\VAN}[3]{#2}
\let\VANthebibliography\thebibliography
\def\thebibliography{\DeclareRobustCommand{\VAN}[3]{##3}\VANthebibliography}

\usepackage{graphicx}	
\usepackage{amsmath}	
\usepackage{rotating}
\usepackage{booktabs}
\usepackage{array}
\usepackage{comment}

\newcommand{\pcm}{\,pc\,cm$^{-3}$}	
\newcommand{\ri}{FRB20121102A}
\newcommand{\re}{FRB20180916B}
\newcommand\T{\rule{0pt}{2.6ex}}       
\newcommand\B{\rule[-1.2ex]{0pt}{0pt}} 

\title[First FRB Detection with the Northern Cross]{The Northern Cross Fast Radio Burst project - II. Monitoring of repeating FRB 20180916B, 20181030A, 20200120E and 20201124A}

\author[M. Trudu et al.]{
M.~Trudu$^{1,2}$\thanks{E-mail: matteo.trudu@inaf.it},
M.~Pilia$^{2}$,
G. Bernardi$^{3,4,5}$,
A.~Addis$^{6}$,
G.~Bianchi$^{3}$, 
A.~Magro$^{7}$,
G.~Naldi$^{3}$, 
D.~Pelliciari$^{3,8}$,
\newauthor
G.~Pupillo$^{3}$,
G.~Setti$^{3,8}$, 
C.~Bortolotti$^{3}$, 
C.~Casentini$^{9,10}$, 
D.~Dallacasa$^{3,8}$, 
V.~Gajjar$^{11}$,
N.~Locatelli$^{12}$,
R.~Lulli$^{3}$,
\newauthor
G.~Maccaferri$^{3}$,  
A.~Mattana$^{3}$, 
D.~Michilli$^{13,14}$, 
F.~Perini$^{3}$,
A.~Possenti$^{1,2}$,
M.~Roma$^{3}$, 
M.~Schiaffino$^{3}$, 
\newauthor
M.~Tavani$^{9,15}$ 
and F.~Verrecchia$^{16,17}$
\\
$^{1}$Università degli Studi di Cagliari, Dipartimento di Fisica, SP Monserrato-Sestu km 0.7, I-09042 Monserrato  (CA), Italy\\
$^{2}$INAF-Osservatorio Astronomico di Cagliari, via della Scienza 5, I-09047, Selargius (CA), Italy\\
$^{3}$INAF-Istituto di Radio Astronomia, via Gobetti 101, 40129 Bologna, Italy\\
$^{4}$South African Radio Astronomy Observatory, Black River Park, 2 Fir Street, Observatory, Cape Town, 7925, South Africa\\
$^{5}$Department of Physics and Electronics, Rhodes University, PO Box 94, Makhanda, 6140, South Africa\\
$^{6}$INAF-Osservatorio di Astrofisica e Scienza dello Spazio di Bologna, Via Piero Gobetti, 93/3, 40129, Bologna, Italy\\
$^{7}$Institute of Space Sciences and Astronomy (ISSA), University of Malta, Msida MSD 2080, Malta\\
$^{8}$Dipartimento di Fisica e Astronomia, Universit\'{a} di Bologna, Via Gobetti 93/2, 40129 Bologna, Italy\\
$^{9}$INAF/IAPS, via del Fosso del Cavaliere 100, I-00133 Roma (RM), Italy\\
$^{10}$INFN Sezione di Roma 2, via della Ricerca Scientifica 1, I-00133 Roma (RM), Italy\\
$^{11}$Department of Astronomy,  University of California Berkeley, Berkeley CA 94720\\
$^{12}$Max-Planck-Institut f\"ur Extraterrestrische Physik (MPE), Giessenbachstrasse 1, 85748 Garching bei M\"unchen, Germany\\
$^{13}$MIT Kavli Institute for Astrophysics and Space Research, Massachusetts Institute of Technology, 77 Massachusetts Ave, Cambridge, MA 02139, USA\\
$^{14}$Department of Physics, Massachusetts Institute of Technology, 77 Massachusetts Ave, Cambridge, MA 02139, USA \\
$^{15}$Università degli Studi di Roma "Tor Vergata", via della Ricerca Scientifica 1, I-00133 Roma (RM), Italy\\
$^{16}$SSDC/ASI, via del Politecnico snc, I-00133 Roma (RM), Italy\\
$^{17}$INAF-Osservatorio Astronomico di Roma, via Frascati 33, 00078 Monte Porzio Catone (RM), Italy}
\date{Accepted XXX. Received YYY; in original form ZZZ}

\pubyear{2021}

\begin{document}
\label{firstpage}
\pagerange{\pageref{firstpage}--\pageref{lastpage}}
\maketitle

\begin{abstract}
In this work we report the results of a nineteen-month Fast Radio Burst observational campaign carried out with the North-South arm of the Medicina Northern Cross radio telescope at 408~MHz in which we monitored four repeating sources: FRB20180916B, FRB20181030A, FRB20200120E and FRB20201124A. We present the current state of the instrument and the detection and characterisation of three bursts from FRB20180916B. Given our observing time, our detections are consistent with the event number we expect from the known burst rate ($2.7 \pm 1.9$ above our 10$\sigma$, 38~Jy~ms detection threshold) in the 5.2 day active window of the source, further confirming the source periodicity.
We detect no bursts from the other sources. We turn this result into a 95\% confidence level lower limit on the slope of the differential fluence distribution $\alpha$ to be $\alpha > 2.1$ and $\alpha > 2.2$ for FRB20181030A and FRB20200120E respectively.
Given the known rate for FRB20201124A, we expect $1.0 \pm 1.1$ bursts from our campaign, consistent with our non-detection.
\end{abstract}




\section{Introduction}
\label{sec:intro}

Fast Radio Bursts (FRBs) are millisecond-duration radio transients with high fluences ($\sim 1-100$\,Jy\,ms) and (mostly) extragalactic origin \citep{petroff2019review, cordes2019review, petroff2021review}.
The discovery of repeating FRBs (or repeaters, \citealt{spitler16,chime_rep19,chime8_19}) has opened the window to in-depth studies of some of these sources, which have provided the most stringent constraints to FRB models so far.
Not all repeaters seems to behave similarly and possess the same kind of progenitor. 

The first known repeater, \ri\footnote{FRB sources are named according to the Transient Name Server (\url{https://www.wis-tns.org/})} \citep{spitler14,spitler16} appears to be a very active and, possibly, very young source, located in the star forming region of a dwarf galaxy at $z = 0.193$ \citep{chatterjee17,tendulkar17,marcote17}, corresponding to a luminosity distance of $\sim 1$\,Gpc.
The presence of a persistent radio source co-located with the FRB emission \citep{marcote17}, the observation of significant variations in dispersion measure (DM) and rotation measure \citep{Michilli18}, and the high number of detections at high frequencies (up to 8~GHz, \citealt{gajjar18,zhang18}) compared to a low number (only one) of detections at low frequencies (below 1~GHz, \citealt{josephy19}), all concurred to interpret this source as a very young compact object surrounded by a dense medium. In particular, many of the models relied on an active magnetar, as the progenitor of \ri, possibly residing in its wind nebula. \citep[see ][]{zhang20Nat}.

The third discovered repeater, \re\ \citep{chime8_19}, was also soon found to be a very active source, which was localised to the outskirts of a star forming region in a nearby massive spiral galaxy ($z=0.0337$, \citealt{marcote20,tendulkar21}). A high activity rate is the main common feature between the two sources.  \re\ is not coincident with a persistent source, with an upper-limit on the luminosity that is forty times lower than the persistent source associated with \ri; no significant DM or RM variations have been observed in time and its emission seems prominent at low frequencies (below 1 GHz, see, e.g., \citealt{marcote20, pilia20,chawla20,pleunis21a,pastor-marazuela21}) while it has never been observed above 2 GHz \citep{pearlman20}.
\re\ was also the first repeater for which a periodicity was established \citep{chime_period_20}, which afterwards led to a similar finding for \ri\ \citep{rajwade20}. \re\ has a period of $16.33$\,days \citep[refined by][]{pleunis21a} with an active window of $\pm 2.6$\,days around its peak phase. \ri, on the other hand, has a period of 161\,days \citep[refined by][]{cruces21} with a 54\% duty cycle.  
The discovery of periodicity, in particular in the case of the nearby and frequently active \re\ has made extensive sensitive follow-up possible and rewarding, allowing for an unprecedented availability of radio data on this source. 

The study of FRBs has received significant momentum from the advent of the CHIME telescope \citep{chime} operating as a transit instrument and being able to monitor the transient sky virtually full time and with real time capability to analyse the data and trigger alerts.
The success of the CHIME/FRB experiment, which led in one year to the discovery of $\sim 500$ FRBs, about 20 of which being repeating sources \citep{chime_cat1_21}, demonstrated that field of view and time on sky are two strong requirements to carry out extensive FRB searches.
The experience that has built up on that, and on the continuous study of repeaters in particular, has further highlighted the need of an immediate reaction on possible burst alerts in order to both understand multi-frequency or chromatic properties of the radio emission and try to catch the elusive multi-wavelength counterpart of this emission.

The Northern Cross (NC) radio telescope, located in Medicina near Bologna (Italy), is a transit telescope which operates at 408 MHz (P-band) with an observational bandwidth of 16 MHz. It is a T-shaped interferometer with two arms aligned along the North-South and East-West directions. The North-South arm has been going through a software and hardware upgrade which made it suitable for FRB observations, whereas the East-West arm is not currently in use. The system and its survey capabilities are described in \cite{locatelli20}, hereafter Paper I. 

In this paper we present the results of an observational campaign, spread over about nineteen months, which monitored four repeaters, mainly performed during known or presumed active phases in the NC observing band. The selected targets are \re, FRB20181030A, FRB20200120E and FRB20201124A. The paper is organised as follows: in \S\ \ref{sec:system} we report the current state of system deployed at the NC for the FRB data acquisition and the FRB detection pipeline; in \S\ \ref{sec:targets} we describe the targets selected in this observational campaign; in \S\ \ref{sec:bursts}  we report the results obtained from this monitoring and in \S\ \ref{sec:conclusion} we provide a short summary.

\section{System description}
\label{sec:system}

\subsection{Data Acquisition System}
\begin{figure}
	\includegraphics[width=\columnwidth]{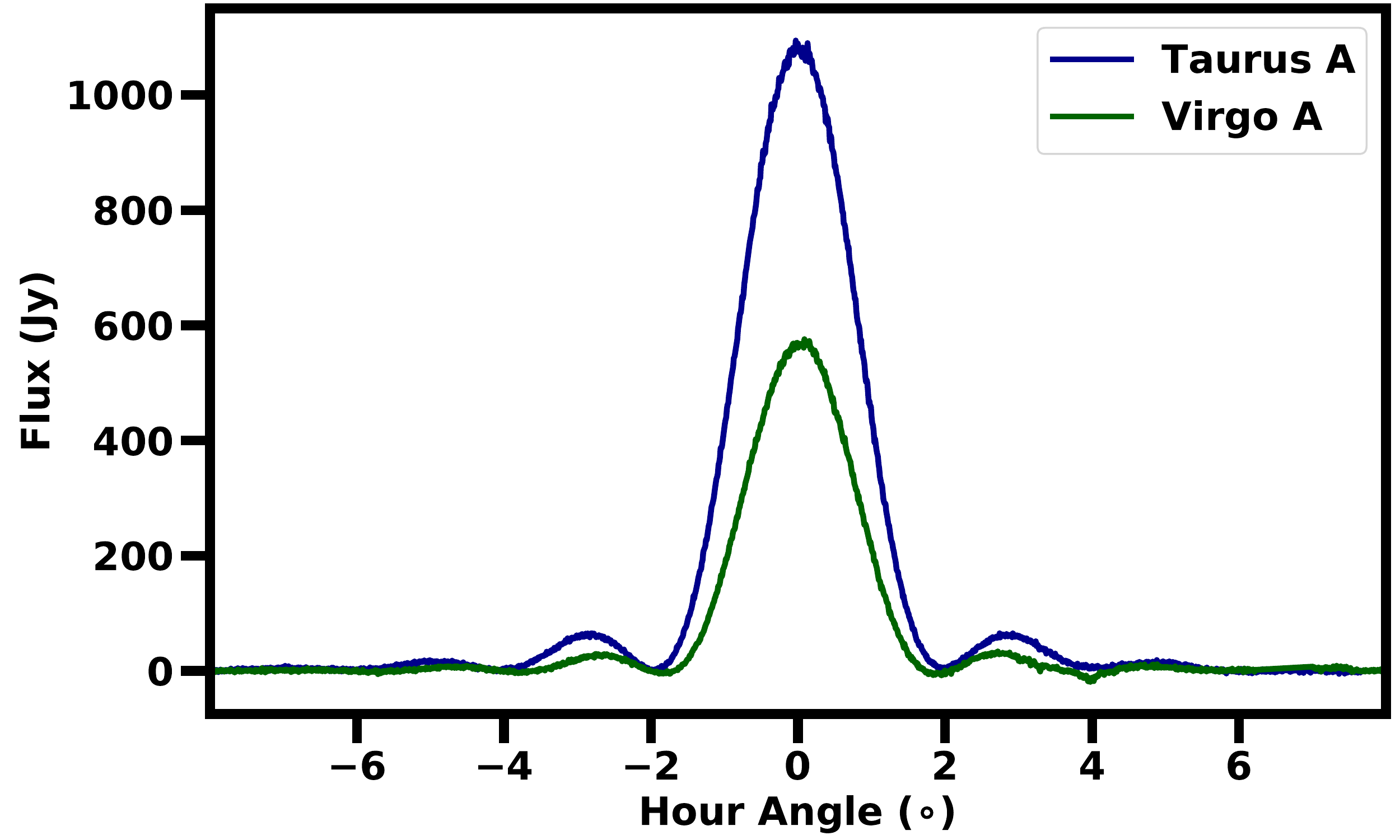}
    \caption{Calibrated drift scans of the astronomical sources, Taurus A and Virgo A, used for the SEFD estimation, observed with the eight-cylinders array beam, referred to the channel corresponding at the central frequency of 408.45 MHz. The profile clearly shows the main and the first lobe of the array beam.}
   \label{fig:sourcetransit}
\end{figure}

\begin{table*}
	\centering
	\caption{Astronomical sources used as calibrators. SEFDs are estimated for groups of 16 dipoles (i.e. one receiver). The SEFD for six and eight cylinders can be obtained by multiplying for the corresponding number of receivers, 24 and 32 respectively.}	
	\label{tab:calibrators}
	\begin{tabular}{lccc} 
		\hline
		\hline
		Source & Sky Position & Starting Time & SEFD \T \\
		       & (RA J2000, DEC J2000) & UT & (Jy)\T\B \\
		\hline
		\hline
		Taurus~A   & $05^h34^m31.940^s$,  $+22$°$00'52.20''$ & 2021/04/13 14:21:03 & $9000 \pm 400$ \T \\
		Virgo~A    & $12^h30^m49.423^s$,  $+12$°$23'28.05''$ & 2021/04/01 22:03:02 & $7800 \pm 180$ \T \B \\
		\hline
	\end{tabular}
\end{table*}

In this section we provide a brief overview of the NC system as presented in Paper I and describe the most recent updates. 
The North-South arm of the NC used in our observations includes sixty four parabolic cylinders. Each cylinder illuminates four groups of sixteen dipoles each, for a total of sixty four dipoles. Signals from each of the sixteen dipoles are combined together analogically and fed to a single receiver. Until March 21\textsuperscript{st} 2021, six cylinders were used, for a total of twenty four receivers, whereas afterwards eight cylinders were equipped and used, for a total of thirty two receivers. Signals from  the receivers were digitised, channelised, combined into a single beam and then written to disk. Unlike the system used in Paper I, we have routinely implemented a second channelisation stage with a windowed FFT that effectively leads to a power stream with a $138.24$~$\mu$s time resolution and a 14.468~kHz frequency resolution, in order to reduce the intra-channel smearing for high DM events. The oversampling polyphase filterbank architecture of the first stage channeliser used in Paper I \citep[see][for details]{Comoretto2017_tpm_fw} causes an overlap between adjacent coarse channels by a factor of 5/32 (oversampling of 32/27).
The overlapping portions of each pair of adjacent channels, that coincide with the filter transition region in the channel edges, are discarded and the resulting bandpass, consisting of 1024 fine spectral channels, is seamless and flat.
The 32 bit time series of each frequency channel are then equalised and rescaled, over discrete time intervals of 10~s, in order to correctly represent 6$\sigma$ samples with 16 bit data, using \texttt{digifil} from the {\sc DSPSR} toolkit \citep{dspsr}. Output data are saved to disk using the {\sc SIGPROC} Filterbank file format \citep{sigproc}.

During the observing campaign we used the eight-cylinder system to observe the transit of two bright sources, Taurus~A and Virgo~A, for 2 hours each, in order to further characterise the System Equivalent Flux Density (SEFD), following up on the estimates derived in Paper I from observations of the PSR~B0329+54 pulsar. Figure~\ref{fig:sourcetransit} shows the corresponding transit observations, where the telescope was steered towards the corresponding declination of each source. We perform a standard on-off observation, where we estimate the background contribution as the average of the power away from source, i.e. for hour angles~$> |6^\circ|$, and subtract it from the observed power when the source is within the main beam. We then fit a Gaussian model to the profile full width at half maximum (FWHM) for 21 frequency channels equally spaced across the 16~MHz bandwidth. Taurus~A and Virgo~A are assumed to be 1080~Jy and 569~Jy at 408~MHz respectively \citep{perley17}, and the best fit to the curve peak provids the conversion from counts to Jy for each channel. The SEFD is derived from the rms of the calibrate power away from sources, i.e. for hour angle~$> |6^\circ|$. The SEFD is found to vary up to 20\% across the bandwidth, a negligible variation for the purpose of the current analysis. We eventually average the SEFD estimates to obtain a band-averaged value (Tab.~\ref{tab:calibrators}). 

We note a slight dependence of the SEFD with the Galactic latitude, varying by $\sim 14\%$ from the Galactic plane (Taurus~A is at a $\sim -5^\circ$ Galactic latitude) to high Galactic latitudes (Virgo~A is at a $\sim +74^\circ$ Galactic latitude). This dependence is qualitatively expected as the sky temperature contribution to the SEFD increases towards the plane. At the same time, our results indicate that the sky temperature contribution to the SEFD is minor. For the purpose of this work we will adopt the same SEFD$^*$ estimate for all FRB source, obtained by averaging the Taurus~A and Virgo~A values: 
\begin{equation}
    \label{eq:avgsefd}
    \text{SEFD}^* = 8400 \pm 420 \ \text{Jy} \ .
\end{equation}

\subsection{Single Pulse Search Pipeline}

The search for FRB candidates is currently performed through an adaptation of the {\sc SPANDAK} pipeline \citep{gajjar18}. The pipeline uses {\sc rfifind} from {\sc PRESTO}
\citep{ransom2002} to prepare the radio frequency interference (RFI) mask, which is then used by {\sc Heimdall} \citep{bbb+12} in order to flag out the noisiest frequency channels. The data are then searched by {\sc SPANDAK} through {\sc Heimdall} across a DM range from 0 to 1000 pc cm$^{-3}$ with a signal-to-noise ratio (S/N) loss tolerance in each DM trial of 10\% and the dedispersed time series have been convolved with a maximum boxcar of 1024 samples. 

Candidates found by {\sc SPANDAK} are then further selected according to the S/N, the DM, the width of the burst $\Delta t $, the maximum number $N_w$ of allowed candidates found within a time window $w$ centred at the time of the candidate and the minimum number $N_m$ of distinct boxcar/DM trials (members) clustered into a candidate. Candidates which agree with the following criteria are classified as plausible FRB candidates:
\begin{equation}
\begin{aligned}
\text{S/N} & \geq   10  \ ; \\
\text{DM} & \geq    10 \ \text{pc} \ \text{cm}^{-3} \ ; \\
\Delta t  & \leq   141.6 \ \text{ms} \ ; \\
N_{w = 4 \ \text{s}}     & \leq   2  \ ; \\
N_m        & \geq   10 \ .  
\end{aligned}
\end{equation}
Filtered candidates are then validated by the artificial neural network classifier {\sc FETCH} \citep{agarwal2020} and, eventually, manually inspected.

\section{Selected Targets}
\label{sec:targets}

We report the results of the observational campaign conducted between the 16\textsuperscript{th} of January 2020 and the 29\textsuperscript{th} of August 2021 for the following four repeating FRB sources: FRB20180916B, FRB20181030A, FRB20200120E, FRB20201124A. 
Sources were selected for their proximity (as initially suggested by their low DM values and later confirmed by their localisation) and therefore as favourable targets both for our new system and for the multi-wavelength campaign that included the NC \citep[see, e.g.][]{pilia20, tavani20, tavani20b}.

\begin{figure*}
	\includegraphics[width=\linewidth]{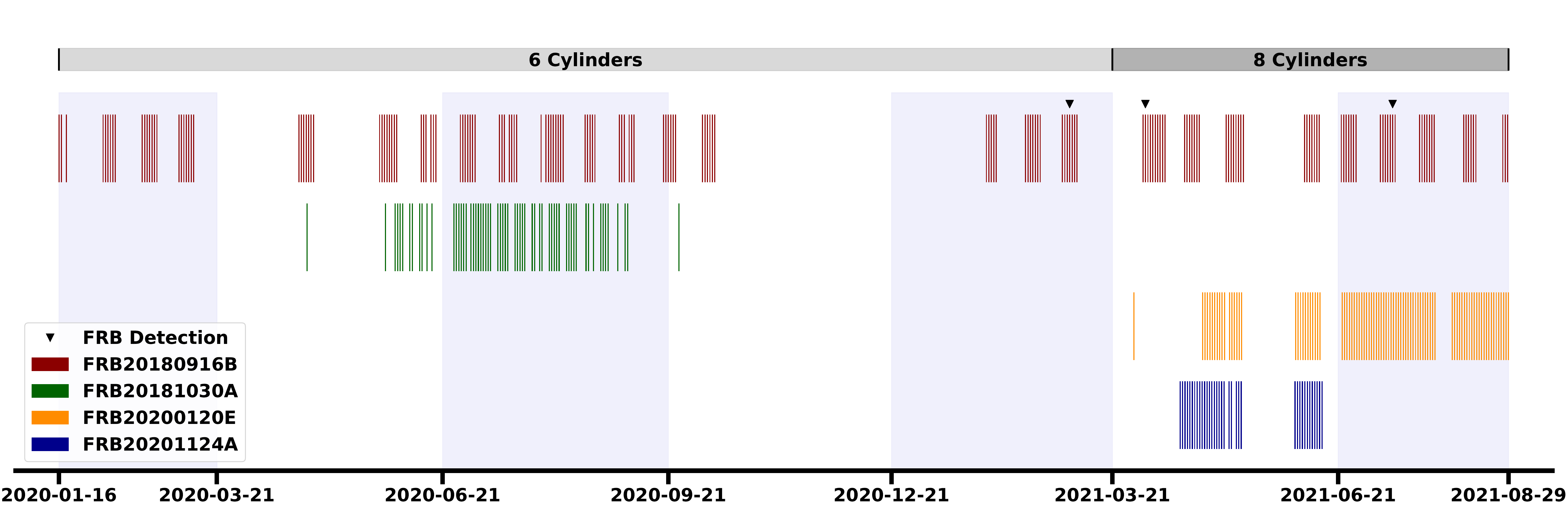}
    \caption{Observational campaign of the NC telescope. Coloured bars represent observations performed for each FRB source as a function of the day of the year. Black triangles indicate days when FRB were detected. Seasons of the year have been depicted as alternated white and lavender rectangles, to help the eye. The top grey and dark grey bars represent, respectively, the time in which the six-cylinders and eight-cylinders system were used during the campaign.}
   \label{fig:campaign}
\end{figure*}

The whole NC campaign is summarised in Fig.~\ref{fig:campaign}, including all the observations performed for the monitored sources at the various epochs. We highlight the time when the transition between the six-cylinders system and the eight-cylinders system occurred.
In the following sections we will describe the targets in more detail.

\subsection{FRB20180916B}

FRB20180916B is our main target and was observed for a total of $\sim 180$~hours throughout the campaign.
Starting in January 2020, when its periodic activity was announced \citep{chime_period_20}, the NC observed FRB20180916B regularly during its active cycles. FRB20180916B has a periodic activity of $\sim 16$~days with an active window of 5.2 days and we observed the source for about seven days each cycle, beginning one day before the period (predicted in the CHIME/FRB bandwidth) in order to match the multi-wavelength campaign from the Swift and AGILE satellites, trying to catch earlier emission \citep{casentini2020,tavani20b, verrecchia21}.
The primary aim of these observations was indeed to find theoretically predicted multi-wavelength counterparts \citep{lyubarsky14,beloborodov17,kumar17,ghisellini18,metzger2019,lkz20,lyutikov20}, looking for time coincidences with other instruments \citep[see][for an updated review]{nicastro21}.

\subsection{FRB20181030A}

FRB20181030A is the fourth known repeater, with two bursts detected by CHIME/FRB in October 2018 \citep{chime8frb} and seven new bursts detected in January 2020. It has a DM~$\sim 103$~\pcm~and a maximum estimated redshift of $z = 0.05$. The star-forming spiral galaxy NGC~3252 ($z \sim$ 0.004) has been identified as its most auspicious host among seven plausible galaxies within the 90\% confidence localisation region \citep{bhardwaj2021}. 

Due to its small DM value, with an associated distance of $\sim 20$~Mpc, FRB20181030A has been monitored, on an approximately daily basis, from April 2020 until September 2020, for a total of 93 hours, as an interesting target for multi-wavelength observations despite the initial lack of localisation. This source, however, has shown very little activity over the last years, compared to FRB20180916B: only nine bursts were detected by CHIME since its discovery, with a fluence smaller than 10~Jy~ms.

\subsection{FRB20200120E}

FRB20200120E is a repeater with DM $\sim$ 88 \pcm \citep{bhardwaj2021a}, initially localised in the outskirts of M81, a spiral galaxy with a distance of $\sim$~3.6~Mpc \citep{m81paper} and afterwards precisely localised in a globular cluster within M81 with the detection of five bursts from the source at 1.4 GHz (L-band) by the European VLBI Network (EVN) \citep{kirsten2021}. Thanks to the relative proximity of the source and also thanks to its high Galactic latitude ($\sim 41.2^\circ$), which makes  the scattering broadening due to the Milky Way interstellar medium negligible, an ultra-high-time resolution analysis of the five bursts detected by \cite{kirsten2021} has been performed, showing that this source can produce nanosecond duration isolated bursts with 10\textsuperscript{41}~K brightness temperature \citep{nimmo2021}, similar to the Crab pulses \citep{hankins03}. This unprecedented finding marked a bridge between young Galactic pulsars and magnetars and the more distant FRBs in terms of burst durations and luminosities \citep[see in particular Fig. 3 from][]{nimmo2021}.

This target was included in the selection being the closest known repeater so far. Analogously to FRB20181030A we monitored this source about once a day from March 2021 to the last day of the campaign reported in this paper, for a total observing time of $\sim 109$~hours.

\subsection{FRB20201124A}

FRB20201124A is a repeater with DM~$\sim 410$~\pcm, discovered by CHIME/FRB in November 2020. It had a very active phase between March and May 2021 \citep{2021ATel144971C, lanman2021}, with a plethora of follow-up detections by other radio telescopes \citep{2021ATel145021K, 2021ATel146031M, 2021ATel145181X, 2021ATel145261L, 2021ATel145381W} at both P and L bands, with an initial localisation of the source by ASKAP \citep{2021ATel145151D}, FAST \citep{2021ATel145181X}, uGMRT \citep{2021ATel145381W} and VLA \citep{2021ATel145261L} and further refined, with a milliarcsecond precision, by the EVN \citep{nimmo2022}. It was localised in a nearby ($ z \sim$ 0.098) \citep{fong2021} galaxy with a high star formation rate which suggests a new-born magnetar as the most likely progenitor \citep{piro21}. We  monitored this source daily from April 2021 to June 2021 for a total observing time of 68 hours.

\section{Bursts Detected}
\label{sec:bursts}

\begin{figure*}
	\includegraphics[width=\linewidth]{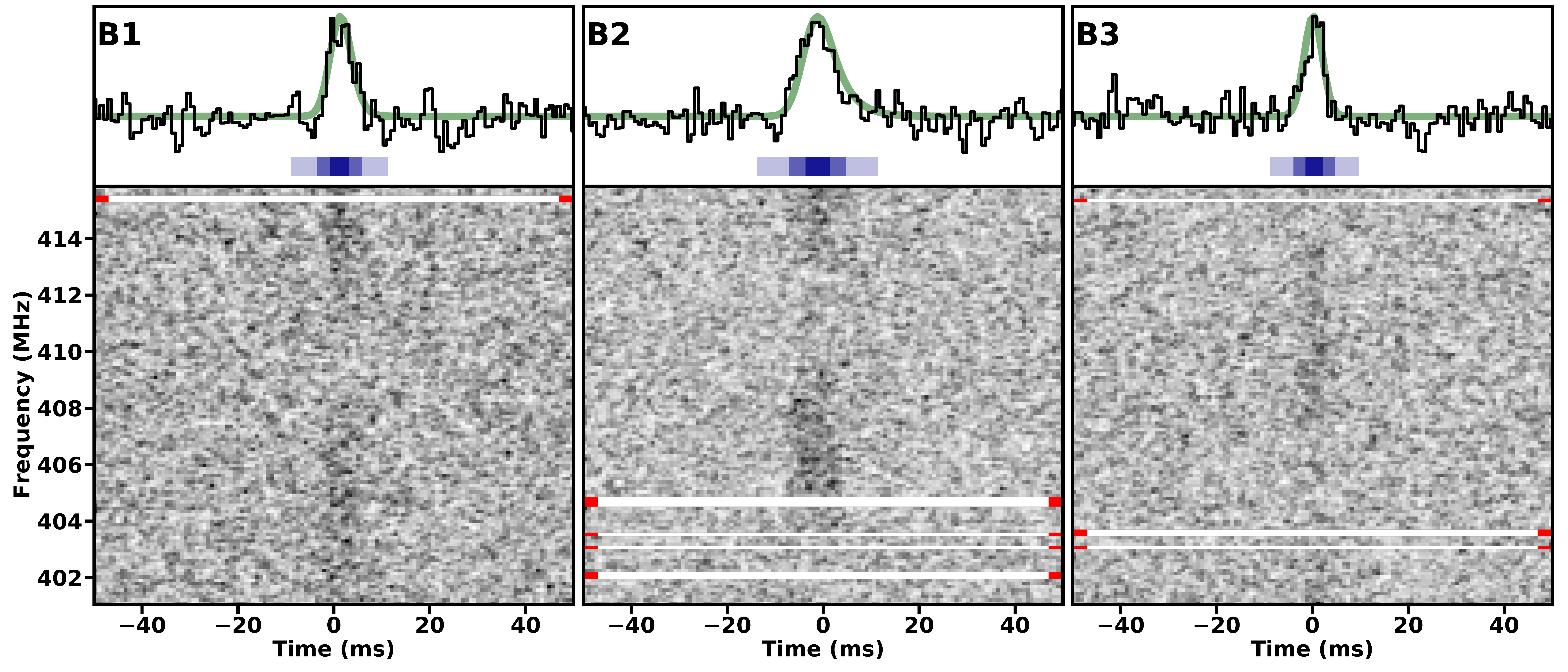}
    \caption{Bursts observed from FRB20180916B (B1, B2 and B3) during the NC observational campaign. Bottom-(left,centre and right) panels represent the dynamic spectrum of the signal, whereas the top-(left, centre and right) are the frequency averaged time series (black curves) and the best-fit model (green curves) of the three bursts. For each burst the 2$\sigma$, FWHM and 10$\sigma$ widths are displayed as blue rectangles. The bursts are, from left to right, ordered by time and each of them has been dedispersed according to the obtained best DM (see Tab. \ref{tab:burstproperties}). For a better display, data are down-sampled to have 128 frequency channels with a 0.116 MHz width each and 128 time bins with size of 0.78 ms. Horizontal white rows (highlighted with red ticks) are flagged channels due to RFI.}
   \label{fig:ncbursts}
\end{figure*}

We report the detection of three bursts from FRB20180916B: B1, B2 and B3, from now onward. These bursts happened on March 3\textsuperscript{rd} 2021, April 3\textsuperscript{rd} 2021 and July 13\textsuperscript{th} 2021, respectively. Tab.~\ref{tab:ncbursts} contains the observed properties of the detected bursts. Fig.~\ref{fig:ncbursts} shows the dedispersed waterfall plots of B1, B2 and B3, obtained with the fit-optimised DM reported in Tab.~\ref{tab:ncbursts}.

\subsection{Burst Characterisation}
\label{sec:burst_characterisation}


The properties of the detected bursts, that is their time of arrival (TOA), the width $\Delta t$, the best DM and the scattering time $\tau$ have been computed by making a fit of the spectro-temporal data array. We use, as a template for the burst in the time domain, a Gaussian function convolved with an exponential decay function \citep{mckinnon20114} and a Gaussian function for the burst in the frequency domain. The fit procedure has been performed using the software package {\sc burstfit}, a detailed description of this package can be found in \cite{aggarwal2021r1comprehensive}.

The flux density $S$ of the incoming radiation from the source is  then computed by using a modified version of the standard radiometer equation for single pulses \citep{handbookpulsar}: 

\begin{equation}
    \label{eq:radiometer}
     S = \text{S/N} \ 
    \frac{\text{SEFD*}}{A \sqrt{ N_p N_c (1 - \xi) \Delta \nu_{\rm ch} \Delta t }}
    \ \zeta \left(\text{TOA} \right)\ .
\end{equation}
Here S/N is the integrated signal-to-noise ratio of the frequency averaged time series, $A$ = 24 or 32 is a geometric factor which takes into account the ratio between the collecting area of the six or eight cylinders system and one receiver (see Paper I for further details), $N_{p}$ = 1 is the number of polarisations, $N_{c}$ = 1024 is the number of spectral channels of the observation, $\xi$ is the fraction of channels excised as RFI and
$\Delta \nu_{\rm ch}$ is the channel width. The multiplicative factor $\zeta \left(\text{TOA}\right)$ takes into account the primary beam attenuation at the burst TOA.

The estimated fluences $F$ of the bursts were calculated as the product between the flux density $S$ and the duration of the burst $\Delta t$.

\begin{table*}
	\centering
	\caption{Properties of the detected bursts of FRB20180916B from the NC campaign. The second column reports the percentage of channels excised as RFI; the third, the fourth and the fifth columns report the barycentric ($\infty$ MHz) time of arrival of the bursts as MJD, UT and phase of the activity period of FRB20180916B (see \S\ \ref{sec:bursts_properties}); the sixth column reports the fit-optimised DM; the seventh column the S/N; the eighth column reports the FWHM burst duration in ms; the ninth column reports the scattering time computed with respect to the reference frequency of 408 MHz; the tenth and eleventh columns report respectively the flux densities and the fluences of the bursts. 
	}
	\label{tab:ncbursts}

	\label{tab:burstproperties}

 \resizebox{\textwidth}{!}{%
	\begin{tabular}{lcccccccccc} 
		\hline
		\hline
		& 
		$\xi$ &
		TOA      &  
		TOA    &  
		$\phi$ &
		DM  & 
		S/N & 
		$\Delta t$ & 
		$\tau$&
		$S$ & 
		$F$ \T \\	
		        & (\%)  & (MJD) &   (UT)  & & (pc cm$^{-3}$) &  &  (ms) & (ms) & (Jy) & (Jy ms) \T\B \\
		\hline
		\hline
        B1 & 2  & 59276.5954859605(4) & 2021-03-03 14:17:29.987(1) & 0.554$\pm$ 0.008 & 349.28$^{+ 0.25}_{- 0.26}$
        & 14.5 &4.76$^{+ 0.57}_{- 0.63}$  &  / &  20$\pm$2 & 96$\pm$14  \T \\
        B2 & 12 & 59307.5148011862(7) & 2021-04-03 12:21:18.822(4) & 0.447$\pm$0.007 & 349.57$^{+ 0.36}_{- 0.36}$ & 21.7 &5.95$^{+ 0.75}_{- 0.68}$ &  <3.6  & 22$\pm$3 & 135$\pm$19  \T \\
        B3 & 8 & 59408.2528584486(4) & 2021-07-13 06:04:06.970(6) & 0.616$\pm$0.009 & 349.64$^{+ 0.35}_{- 0.37}$ & 12.5 &4.37$^{+ 0.45}_{- 0.45}$  &  / & 16$\pm$1 & 71$\pm$8 \T\B \\
		\hline
	\end{tabular}

	}%
	\\

\end{table*}

\subsection{Bursts Properties}
\label{sec:bursts_properties}
The top panels of Fig. \ref{fig:ncbursts} show the frequency averaged time series of B1, B2 and B3. We obtain a significant measurement for the scattering time only for B2, with a value of 3.6 ms at 408 MHz. This would correspond to $\sim$ 0.8 ms at 600~MHz, consistent with previous scattering time measurements reported for this source \citep{chime8frb}. However, we consider this value as an upper limit as \cite{marcote20} and \cite{pastor-marazuela21} placed a tighter constraint on the scattering time scale of the order of 3 $\mu$s at 1.7 GHz, similar to the NE2001 \citep{ne2001} prediction of 2 $\mu$s. Our estimate would correspond to 10 $\mu$s at 1.7~GHz. Hence, we conclude that this apparent scattering tail, as showed in the model for B2 in Fig. \ref{fig:ncbursts}, could be originated by the presence of not resolvable sub-bursts \citep{chime8frb}.


None of the three  detected burst show peculiar spectro-temporal features (bottom panels of Fig. \ref{fig:ncbursts}), that is the typical downward drift of the signal in the time-frequency plane (oftentimes called "sad trombone effect") as often reported for repeater sources \citep{chime_rep19, hessels19, pleunis21a}, as can be seen from the dynamic spectra in Fig. \ref{fig:ncbursts}. 

\begin{figure}
	\includegraphics[width=\linewidth]{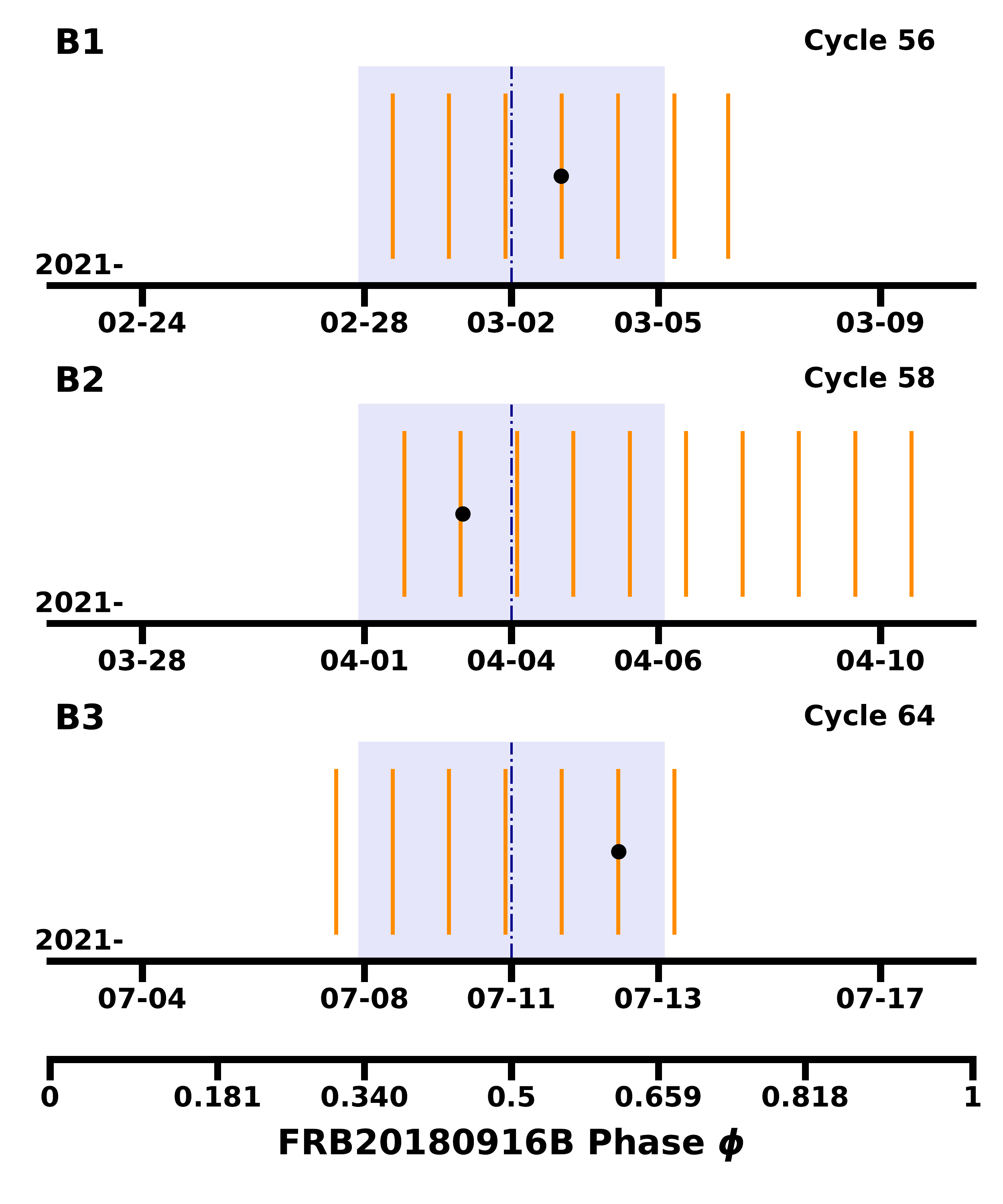}
    \caption{Bursts B1, B2 and B3 as a function of the periodicity phase $\phi$. Orange bars show the NC observations during the source activity cycles (lavender rectangle). Cycle 1 corresponds to the first burst detection \citep{chime8frb}, with a starting epoch $\phi_{0}$ = 58369.40 MJD and $\phi$ = 0.5 to be the centre of the activity window. Our detections (black circles) happened during Cycles 56, 58 and 64 respectively. }
   \label{fig:ncwindows}
\end{figure}

Figure~\ref{fig:ncwindows} shows the span of our observations and the occurrence of B1, B2 and B3 as a function of the relative phase $\phi$, during the activity cycles of FRB20180916B. The phases are obtained folding the data at the nominal period of 16.33 days taking a starting phase $\phi_0$ = 58369.40 MJD (corresponding to Cycle 1), such that  $\phi = 0.5$ corresponds to the peak of the activity of the source \citep[see][for further details]{pleunis21a}.

The obtained phases for B1, B2 and B3 are reported in Tab. \ref{tab:ncbursts}. From Fig. \ref{fig:ncwindows} we see that the three burst are consistently located within the predicted activity window of 5.2 days from CHIME/FRB, since our observational bandwidth overlaps with theirs \citep[see again][Fig. 9]{pleunis21a}. We do not report any detection from outside its window of activity (Fig.~\ref{fig:ncwindows}), consistently with the observed chromatic activity, as burst were detected at $\phi \sim 0.7$ at lower frequencies \citep{pastor-marazuela21, pleunis21a}.


\subsection{Rate Estimation and Comparison with CHIME/FRB}
We estimate the number of expected bursts at our facility for the monitored sources, making a comparison with the detection rates reported by CHIME/FRB, due to the partially overlapping observational bandwidths. Let us assume that the differential number of bursts $dN$, for a facility $x$ which operates at the central frequency $\nu^{x}_c$, with fluence (or equivalently flux density for a 1~ms burst), within the interval $(F , F + dF) $ follows a power law of the kind:
\begin{equation}
    \label{eq:powerlaw}
    \frac{dN^{x}}{dF} = K 
        \left( \frac{\nu^{x}_c}{\nu_{\rm ref}} \right)^{- \beta} 
    \left( \frac{F}{F_{\rm ref}} \right)^{- \alpha}      \ ,
\end{equation}
where $F_{\rm ref}$, $\nu_{\rm ref}$ are respectively a reference fluence and a reference frequency, $K$ corresponds to $dN^{x}/dF$ at $\nu = \nu_{\rm ref}$ and $F = F_{\rm ref}$, $\beta = 1.6 \pm 0.3$ \citep{macquart19} is the spectral index, and lastly $\alpha$ is the slope of the differential fluence distribution \citep[see][\S 6.2]{chime_cat1_21}\footnote{We consider a different convention for the slope $\alpha$. Our $\alpha$ refers to the differential fluence distribution, whereas for the cited reference to the cumulative. The values of $\alpha$ reported in the reference, to be in accord with our convection, should be read as -$\alpha$ + 1.}. The value of $\alpha$ appears to be different for each source, with a value for instance of $\alpha = 2.3 \pm 0.4 $ for FRB20180916B \citep{chime_period_20}, although the possibility of a universal parameter for at least the class of repeaters is still open \citep[see e.g.][for a discussion]{marthi21}. Assuming that an FRB source is observed for an average observing time $\delta t_{\rm obs}^{x}$, we can evaluate the differential burst rate $dR^x/dF$ from Eq. \ref{eq:powerlaw} as:
\begin{equation}
    \label{ratedef}
    \frac{dR^x}{dF} = \frac{d^2N^x}{dFdt} = 
    \frac{K}{\delta t^x_{\text{obs}}}
    \left( \frac{\nu^x_c}{\nu_{\rm ref}} \right)^{- \beta}     
    \left( \frac{F}{F_{\rm ref}} \right)^{- \alpha} \ .
\end{equation}
Henceforth the expected detection rate of bursts with fluence exceeding a given threshold $F^x_l$ is:
\begin{equation}
    \label{eq:rate}
    R^x(F > F^x_l) = R^x = \frac{K}{\delta t^x_\text{obs}}
    \left( \frac{\nu^x_c}{\nu_{\rm ref}} \right)^{- \beta}
    \int_{F^x_l}^{+ \infty}
    \left( \frac{F}{F_{\rm ref}} \right)^{- \alpha} dF \ . 
\end{equation}
Taking $\alpha > 1$, we can ensure the convergence of the integral in Eq. \ref{eq:rate} and we obtain the following expression:
\begin{equation}
    \label{eq:ratefinal}
    R^x= \frac{K}{\delta t^x_\text{obs}}
    \frac{1}{\left( \alpha - 1 \right)}
    \left( \frac{\nu^x_c}{\nu_{\rm ref}} \right)^{- \beta}
    \left( \frac{F^x_l}{F_{\rm ref}} \right)^{- \alpha + 1} \ .
\end{equation}
Considering now Eq.~\ref{eq:ratefinal} for both the NC and CHIME/FRB (CF) and calculating the ratios between the two equations, we can evaluate the rate of bursts expected at the NC with respect to the rate of bursts expected by CHIME/FRB as:
\begin{equation}
    \label{eq:ratenc}
    R^{\text{NC}}= R^{\rm CF}
    \left(\frac{\delta t^{\text{NC}}_\text{obs}}{\delta t^{\rm CF}_\text{obs}} \right)^{-1}
    \left( \frac{\nu^{\text{NC}}_c}{\nu^{\rm CF}_c} \right)^{- \beta}
    \left( \frac{F^{\text{NC}}_l}{F^{\rm CF}_l} \right)^{- \alpha + 1} \ .
\end{equation}
Lastly, the average number of bursts $N^{\text{NC}}$ that we expect at the NC, with fluence greater than $F_l$\textsuperscript{NC}, throughout a campaign of total duration $\Delta T_c$ will be:
\begin{equation}
    \label{eq:avnumb}
    N^{\text{NC}} = 
    R^{\text{NC}} \ 
     \Delta T_c \ .
\end{equation}

\subsubsection{Instrument Fluence Detection Threshold}
\begin{figure}
	\includegraphics[width=\linewidth]{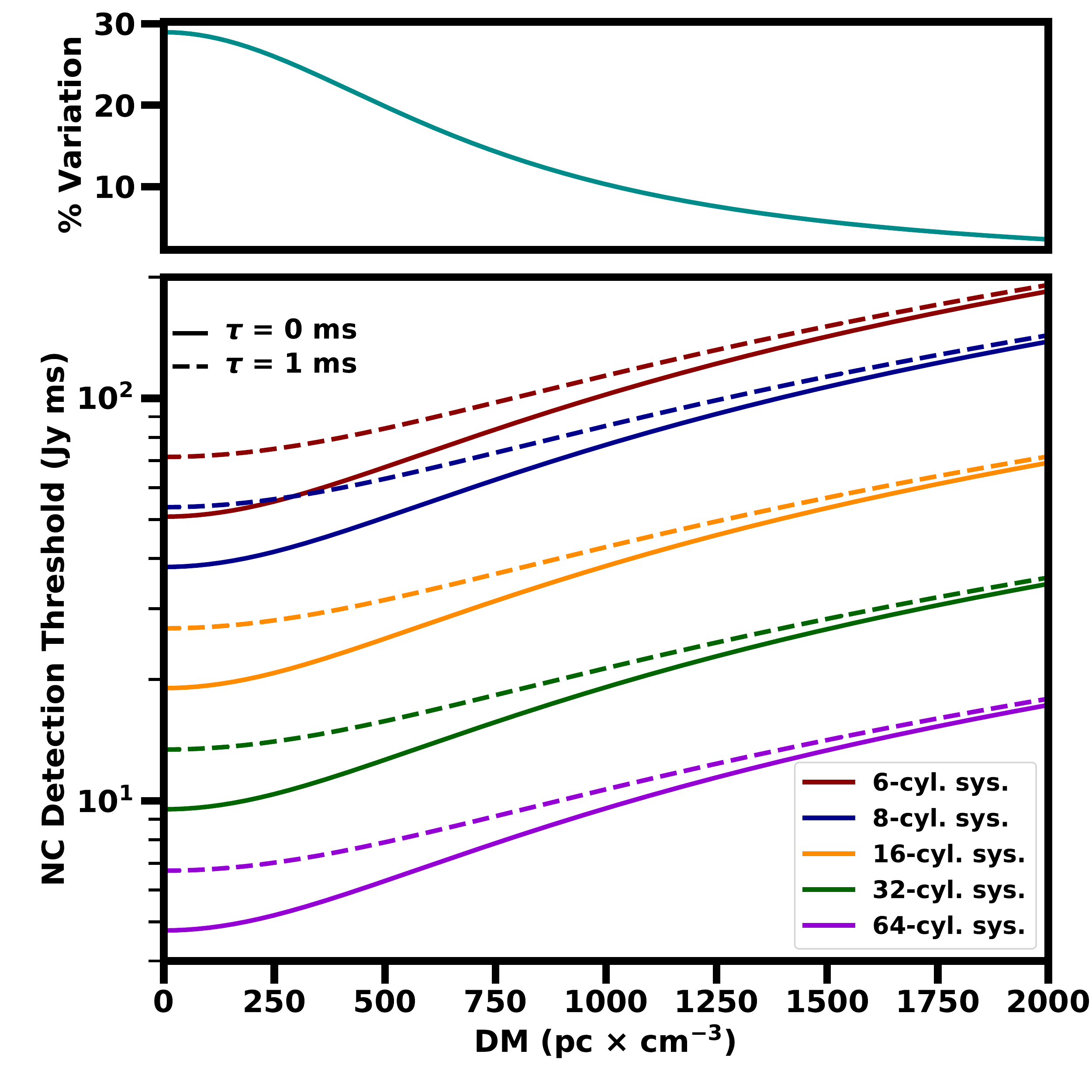}
    \caption{Northern Cross fluence detection threshold curves. The bottom panel shows the NC fluence detection threshold as a function of the DM computed via Eq. \ref{eq:sensitivity} considering a scattering time $\tau = 0$ (solid lines) and 1 ms (dashed lines), for the previous six-cylinders system (red), the current eight-cylinder system (blue) and the future sixteen, thirty two, sixty four-cylinder systems (orange,green,violet). The top panel represents the percentile variation between the threshold curve at $\tau$ = 1 ms and the curve at $\tau$ = 0. }
    \label{fig:ncsensitivity}
\end{figure}

In order to evaluate the number of bursts throughout the campaign we need to estimate the minimum fluence detectable, given a certain threshold, we can achieve with the NC. In general this fluence will depend on the physics which impact the arrived signal (e.g. the scattering) and also on the instrumental performances (e.g. the sampling time). 
From the radiometer equation we can compute the minimum flux density $S'_l$, considering a minimum S/N of 10, we are able to detect with the NC \citep{burkespoalor2011}:
\begin{equation}
    \label{eq:fluxmin1}
    S_l = S'_l \  \frac{\Delta t_m}{\Delta t} \ ,
\end{equation}    
\begin{equation}
    \label{eq:fluxmin2}
    S'_l = 10 \times 
   \frac{\text{SEFD*}}{A \sqrt{N_p N_c  \Delta \nu_{ch} \Delta t }} 
   \ \zeta \left(\text{TOA} \right)  \ ,
\end{equation}
where 
\begin{equation}
    \label{eq:widthobs}
    \Delta t_m = \sqrt{\Delta t^2 + t^2_{\text{DM}} + \tau^2 + t^2_s}
\end{equation}
consists in the measured width of the burst, which will be generally broadened by the scattering time $\tau$, the sampling time $t_s$ of the receiver and the intra-channel smearing:
\begin{equation}
    \label{eq:tdm1}
    t_{\text{DM}} = 8.3 \times 10^{-3} \ \left(\frac{\text{DM}}{\text{pc} \  \text{cm}^{-3}} \right)
    \left(\frac{\Delta \nu_{\rm ch} }{\text{MHz}} \right) 
    \left( \frac{\nu_c}{\text{GHz}} \right)^{-3} \ \text{ms} \ .
\end{equation}
The quantity $S'_l$ in Eq. \ref{eq:fluxmin2} corresponds so to the minimum flux density detectable in the case of negligible intra-channel smearing, scattering and sampling time.
Considering now a nominal width of a burst of $\Delta t = 1$ ms and  assuming $\zeta = 1.5$ throughout the whole transit of the source\footnote{$\zeta$ can actually vary between one and two throughout our observations, following the primary beam variations in a transit observation.}, substituting all the numbers in Eq. \ref{eq:fluxmin1} we can express the minimum detectable fluence (at 10$\sigma$) $F_l^{\rm NC}$ of the instrument as: 
\begin{eqnarray}
    \label{eq:sensitivity}
    F_l^{\text{NC}} &=& S_l \left(\Delta t  = 1 \ {\rm ms} \right) \times 1 {\rm ms} \\
    &=& \frac{690.12} {\text{A}}
    \frac{\sqrt{1.02 +t^2_{\text{DM}} + \tau^2} \, \text{ms}}{1 \, \text{ms}} \ \  \text{Jy ms} \ .
\end{eqnarray}
Figure \ref{fig:ncsensitivity} displays the fluence detection threshold of the NC, computed via Eq. \ref{eq:sensitivity}, as a function of DM. In addition to the previously considered values of 24 and 32 for the six and eight- cylinders systems, we consider A = 64, 128, 256 for the futures sixteen, thirty-two and sixty-four cylinders systems, respectively.

The scattering time is more uncertain to estimate, and we consider two cases: the case in which we neglect it and the case of a scattering time of 1 ms. From Fig. \ref{fig:ncsensitivity} we see the detection threshold increase as the DM increases, consistently with the fact the DM smearing dominates at higher DM. Regarding the scattering time, from the top panel of Fig. \ref{fig:ncsensitivity} we see that it is quite relevant at low DM, whereas for DMs higher than 1000 \pcm~the relative variation between the two defined regimes is below the 10\%.


We estimate the minimum fluence that can be detected from the monitored sources using Eq.~\ref{eq:sensitivity} and assuming $\tau = 0$.
In the case of FRB20180916B, we can detect bursts with fluences greater than 51 and 38 Jy ms for the six and eight cylinders system respectively. 
FRB20181030A has been monitored only when the six-cylinders system were in place and we estimate a fluence detection threshold of 44 Jy ms. FRB20200120E and FRB20201124A were both monitored with the eight-cylinders system and we place for them, respectively, a fluence detection threshold of 33 and 42 Jy ms.   

With current upgrades in progress, when the full North-South arm will be in use, we expect to find bursts, for instance from FRB20180916B, with 5 Jy ms fluence at S/N = 10.

\begin{table}
	\centering
	\caption{Results for the NC observational campaign. For each source we report the observational system deployed (see \S\ \ref{sec:system}) in the second column, the fluence detection threshold in the third column and the total observing time in the fourth column. For FRB20180916B and FRB20201124A we report in the fifth column the expected number of bursts $N^{\rm NC}$, whereas for the other two sources we report the 95\% confidence level lower limits of the slope of the differential fluence distribution $\alpha$.}
	\label{tab:expnumber}

	\begin{tabular}{lcccc} 
		\hline
		\hline
		 Source & System & $F_l^{\rm NC}$ & $\Delta T_c$  & $N^{\rm NC}$  \T \\
	     &  &   (Jy ms)   & (hours) &   \T \B \\
		\hline
		\hline
		FRB20180916B & Six  & 51 & 112.7  & $1.4 \pm 1.5$ \T  \\
	    & Eight & 38 & 70.4   & $1.3 \pm 1.2$ \T \\
	    &  & & 183.1  & $2.7 \pm 1.9$ \T \\
	    FRB20201124A & Eight & 44 & 68.14  &  1.0 $\pm$ 1.1 \T \\
	    & & &  \T \\
	    \hline
	    \hline
	    Source & System & $F_l^{\rm NC}$ & $\Delta T_c$ & $\alpha$  \T \\
	    & & (Jy ms) & (hours) &   \T \B \\
	    \hline 
	    \hline 
	    FRB20181030A & Six   & 33 & 93.0  & >2.1  \T \\
	    FRB20200120E & Eight & 42 & 109.2 & >2.2  \T \B\\
		\hline
	\end{tabular}

\end{table}

\subsubsection{FRB20180916B}


Table \ref{tab:ncbursts} reports the flux densities and the fluences of the three bursts detected, computed from Eq. \ref{eq:radiometer}.
We compute the expected number of bursts from FRB20180916B above our detection threshold by using Eqs. \ref{eq:avnumb} and \ref{eq:ratenc}.
In order to do so, we make the following assumptions. The observing time of CHIME/FRB requires the knowledge of the time when the source was within the FWHM of their beam at 600 MHz: as observing time for CHIME/FRB we assume 70\% of the computed transit time for the source\footnote{Source transit time can be computed thanks to the CHIME/FRB Online Calculator: \url{https://www.chime-frb.ca/astronomytools}}, yielding to 8.4~minutes.
With the NC we observed the source for approximately 66 minutes every day of the campaign. We assume the burst rate to be $0.9 \pm 0.5$~hours\textsuperscript{-1} above a fluence limit of 5.2 Jy ms within its activity window of 5.2 days \citep{chime_period_20}.
In Tab. \ref{tab:expnumber} we report the expected number of bursts for $\sim$ 102 hours (six-cylinders) and $\sim$ 70 hours (eight-cylinders). We expect $1.4 \pm 1.5$ and $1.3 \pm 1.2$ bursts respectively, and a total of $2.7 \pm 1.9$ bursts for the total monitoring of $\sim$ 180 hours, consistent with our three detections. 
When the full North-South arm will be operational, we can expect $\sim$ 36 bursts from this source for a campaign of the same duration as the one performed.



\subsubsection{FRB20181030A, FRB20200120E and FRB20201124A}

\begin{figure}
	\includegraphics[width=\linewidth]{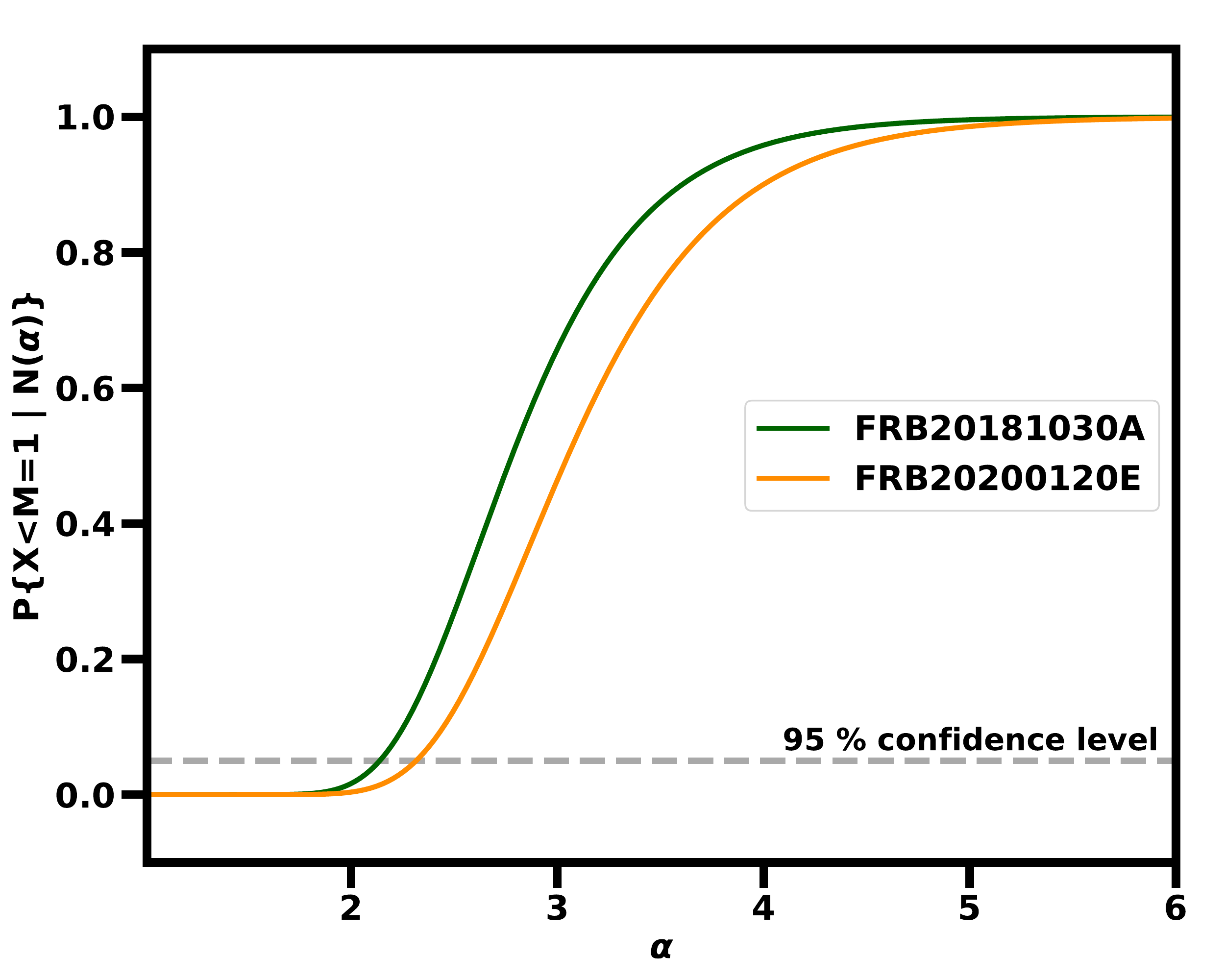}
    \caption{Cumulative Distribution Function of detecting zero events as a function of the slope of the differential fluence distribution $\alpha$, for FRB20181030A (green line) and FRB20200120E (orange line). The horizontal dashed grey line demarcates the  5 \% probability of not finding bursts from FRB20181030A (FRB20200120E) with $\alpha$ less then 2.1 (2.2), implying that $\alpha$ has to be greater than 2.1 (2.2) with a 95 \% confidence level.}
    \label{fig:ratelowerlimit}
\end{figure}

No detections were obtained for FRB20181030A, FRB20200120E and FRB20201124A after an observing campaign of 93, 109 and 68 hours, respectively. We used our observations to constrain $\alpha$, the slope of the fluence distribution for each source (see Eq.~\ref{eq:powerlaw}).
Following \cite{amiri17} and Paper I, if we assume that the occurrence of a burst is a Poissonian process, we can compute the likelihood of detecting $M$ bursts with expectation number $N(\alpha)$ (computing this with Eq. \ref{eq:ratenc} and Eq. \ref{eq:avnumb}) as: 
\begin{equation}
    \label{eq:poission}
    p\{ M; N(\alpha) \} = \frac{N(\alpha)^{M} e^{- N(\alpha)}}{M !} \ .
\end{equation}
The cumulative distribution function (CDF) of seeing $X$ events lower than M , with $N(\alpha)$ expected, results then as the following expression:
\begin{equation}
    \label{eq:cdfm}
    P
    \{ X < M | N (\alpha)  \} = 
    \sum_{k=0}^{M - 1} p \{ k; N(\alpha) \}  \ .
\end{equation}
Hence in the case of less than M = 1 events (non-detection case):
\begin{equation}
    \label{eq:cdf}
    P
    \{ X < M = 1 | N (\alpha)  \} = e^{- N (\alpha)} \ .
\end{equation}
In order to estimate $N(\alpha)$ we make some assumptions for the three sources. We separate the case of FRB20181030A and FRB20200120E from that of FRB20201124A for reasons that we report below.

Figure \ref{fig:ratelowerlimit} shows the CDF for the detection of zero events as a function of the slope $\alpha$, for FRB20181030A and FRB20200120E, considering their respective total observing time (Tab. \ref{tab:expnumber}). In the cases of FRB20181030A and FRB20200120E, we assume a measured rate for CHIME/FRB equivalent to the predicted rate of the facility: $\sim$ 820 sky\textsuperscript{-1} day\textsuperscript{-1} above a threshold of 5 Jy ms at 600 MHz \citep{chime_cat1_21}. Under the same hypotheses as for FRB20180916B, we set 13 min as the average observing time  at CHIME/FRB for FRB20181030A and 10 minutes for FRB20200120E. The average observing times at the NC for the two sources were 90 and 72 minutes respectively.


Setting a confidence level of 95\%, from Fig. \ref{fig:ratelowerlimit} we can rule out the values of $\alpha$ for which the probability computed by Eq. \ref{eq:cdf} is less than 0.05. The lower limits obtained for the values of $\alpha$ for both sources are reported in Tab. \ref{tab:expnumber}. These limits are  consistent with the estimated values of $\alpha$ for a low-DM population as showed by \cite{chime_cat1_21}. In their work they searched for correlations between fluences and DMs among the current population of detected FRBs and  found that 
 the distributions of fluence versus $\alpha$ peak at  two different values: $\alpha \sim$ 2 for FRBs with DM between 100-500 \pcm~and $\alpha \sim$ 2.8 for FRBs with DM > 500 \pcm~(this retains the value of $\alpha = 2.5$, considering the whole sample of bursts, compatible with an Euclidean Universe).



In the case of FRB20201124A, \cite{lanman2021} report a significant increase of the burst rate from the source in the period March-May 2021 with respect to the period between its discovery in November 2020 and March 2021, implying a non-Poissonian distribution of the events.

Due to this non-Poissonianity we only conservatively estimate the expected number of bursts above the estimated threshold for our facility for this source, in order to asses the compatibility with a non-detection. Following \cite{lanman2021}, we consider 4 minutes as the average observing time and a value of $\alpha = 4.5 \pm 2.2$ for CHIME/FRB. For CHIME/FRB we assume a rate of 5.4 hour\textsuperscript{-1} (see Fig. 3 of the aforementioned paper) above a fluence limit of 17 Jy ms. We observed  this source with NC for 120 minutes on average. The expected number of bursts from FRB20201124A for the NC was calculated as $1.0 \pm 1.1$, consistent with a non-detection.

\section{Conclusions}
\label{sec:conclusion}

This work presents the first FRB detections from the Medicina Northern Cross radio telescope, whose North-South arm is currently equipped to carry out FRB observations at 408 MHz with an observational bandwidth of 16 MHz. We performed a nineteen months observational campaign in which we targeted FRB20180916B, FRB20181030A, FRB20200120E and FRB20201124A.

We describe the facility upgrade from the six-cylinder system to the eight-cylinder system. Before the upgrade we report the detection of a single burst from FRB20180916B above a 10$\sigma$ fluence threshold of 51 Jy ms (which also accounts for the intra-channel smearing for our current frequency resolution of 14.468 kHz). After the upgrade we report the detection of two bursts from the same source above a fluence threshold of 38 Jy ms. All bursts were found within the 5.2 day activity window of the source, confirming the source periodicity. Assuming the CHIME/FRB source rate, we expected to detect $2.7 \pm 1.9$ bursts in our campaign, above the aforementioned fluence detection thresholds, consistent with our results.

We report no detections for the other three sources. In the cases of FRB20181030A and FRB20200120E, we constrain the slope of the differential fluence distribution $\alpha$ to be $\alpha > 2.1$ and $\alpha > 2.2$ at the 95\% confidence level, respectively. In the case of FRB20201124A we estimate $1.0 \pm 1.1$ bursts to be observed above a fluence detection threshold of 42 Jy ms, consistent with our non-detection.

\section*{Acknowledgements}

The authors thanks the anonymous referee for the useful comments which significantly improved the quality of this work. 
The Northern Cross radio telescope is a facility of the University of Bologna operated under agreement by the Institute of Radio Astronomy of Bologna (Istituto Nazionale di Astrofisica, INAF). MT gratefully acknowledges INAF for the financial support for his PhD program. NL acknowledges financial support from the European Research Council (ERC) under the European Union’s Horizon 2020 research and innovation program Hot-Milk (grant agreement No. [865637])

\section*{Data Availability}

The data presented in this paper and the software used can be shared upon reasonable request to the corresponding author.

\section*{Software Packages}
\subsection*{Python Packages}
{\sc burstfit}\footnote{\url{https://github.com/thepetabyteproject/burstfit}} \citep{aggarwal2021r1comprehensive}; {\sc Matplotlib} \citep{matplotlib}; {\sc Numpy} \citep{numpy}; {\sc Scipy} \citep{scipy}; {\sc YOUR}\footnote{\url{https://github.com/thepetabyteproject/your}}\citep{your}.
\subsection*{FRB/Pulsar Softwares}
{\sc DSPSR}\footnote{\url{http://dspsr.sourceforge.net/}} \citep{dspsr};
{\sc FETCH}\footnote{\url{https://github.com/devanshkv/fetch}} \citep{agarwal2020};
{\sc Heimdall}\footnote{\url{https://sourceforge.net/projects/heimdall-astro/}} \citep{bbb+12};
{\sc PRESTO}\footnote{\url{https://github.com/scottransom/PRESTO}} \citep{ransom2002}; 
{\sc SIGPROC}\footnote{\url{http://sigproc.sourceforge.net/}}\citep{sigproc};
{\sc SPANDAK}\footnote{\url{https://github.com/gajjarv/PulsarSearch}} \citep{gajjar18}.



\bibliographystyle{mnras}
\bibliography{biblioMT} 








\bsp	
\label{lastpage}
\end{document}